\documentclass[fleqn,twoside]{article}
\usepackage{espcrc2}

\newif\ifpdf
\ifx\pdfoutput\undefined\pdffalse\else\pdfoutput=1\pdftrue\fi
\ifpdf
  \usepackage[backref,citecolor=blue]{hyperref}
  \let\myhref=\href\def\href#1#2{\penalty-20\myhref{#1}{\tt #2}}%
\else%
  \def\href#1#2{{\penalty-20\tt #2}}
\fi

\def\M{{\cal M}}		
\def\D{{\cal D}}
\def\det{\mathop{\rm det}}	
\def\asqtad{{\scshape asqtad}}	
\def\qcdoc{{\scshape qcdoc}}	
\def\fermionsuper:#1_#2{#1_{\mbox{\tiny #2}}} 

\title{Exact 2+1 Flavour RHMC Simulations}

\author{M. A. Clark,\address[MCSD]{School of Physics, 
        The University of Edinburgh, \\ 
        Edinburgh EH9 3JZ, United Kingdom} 
        A. D. Kennedy,\addressmark\ 
        and
	Z. Sroczynski\address{Division of Theoretical Physics,
	  Department of Mathematical Sciences, \\ University of Liverpool,
	  Liverpool L69 3BX, United Kingdom}}
\begin{document}

\begin{abstract}
\noindent We consider the Rational Hybrid Monte Carlo algorithm for performing
exact 2+1 flavour fermion simulations.  The specific cases of {\asqtad} and
domain wall fermions are considered.  We find that in both cases the na\"{\i}ve
performance is similar to conventional hybrid algorithms.
\end{abstract}

\maketitle

\section{Introduction}

Traditionally, ``2+1'' flavour simulations have been performed using the R
algorithm \cite{gottlieb87a}, but an exact algorithm is clearly desirable.
There also exist exact Polynomial Hybrid Monte Carlo algorithms
\cite{forcrand96a,jansen97a}, which can be used for 2+1 simulations: however,
such algorithms are expensive with regard to memory consumption and for the
case of {\asqtad} fermions such an algorithm would be impractical due to the
very expensive force term calculation.  We explore the use of the Rational
Hybrid Monte Carlo (RHMC) algorithm applied to {\asqtad} and domain wall 2+1
simulations.

\section{Rational Hybrid Monte Carlo}

The RHMC algorithm \cite{Clark:2003na} is an exact algorithm which allows the
simulation of theories where the fermionic determinant is raised to an
non-integer power.  As with conventional Hybrid Monte Carlo
(HMC)~\cite{duane87a}, the determinant is replaced by an integral over the
exponential of an action containing bosonic pseudofermion fields. The non-local
fermion matrix is replaced by a rational approximation,
\begin{eqnarray*}
  \det\M^{\alpha} & = & \int \D \bar{\psi}\D\psi
  e^{-\bar{\psi}\M^{-\alpha}\psi}\\
  & \approx & \int \D\bar{\psi}\D\psi
  e^{-\bar{\psi}r(\M)\psi}.
\end{eqnarray*}
Such approximations are cheap to evaluate since they can be written in partial
fraction form, allowing evaluation using a multi-shift solver.

Rational approximations have a great advantage over polynomial approximations
of the same degree, because rational approximations typically have an error
many orders of magnitude smaller.  It is this important feature, which allows
the use of a conventional Metropolis acceptance test (\(\mbox{cost} \propto
V^{5/4}\)) as opposed to a noisy estimator (\(\mbox{cost} \propto V^2\)).

When performing RHMC, we must ensure that the approximations used for the
heatbath and the evaluation of the accept/reject Hamiltonian are exact (to
machine precision), otherwise we would introduce a systematic error into our
simulation.  This requirement is not true for the molecular dynamics (MD)
evolution, where any error introduced is corrected for by the acceptance test.
As this error increases, a decrease in the acceptance probability will be
observed.  Typically this means that we have two approximations when performing
our simulations, a high order approximation which is used for the heatbath and
accept/reject evaluation, and a lower order approximation used for the MD
evolution through phase space.

\section{ASQTAD Simulations}

{\asqtad} simulations are popular at the moment, since they are computationally
very cheap compared to the chiral approaches favoured for theoretical reasons.
As with conventional staggered fermions, {\asqtad} fermions naturally describe
a theory of four degenerate flavours, and we are forced to take the square root
of the fermion matrix to obtain a theory which we may hope describes two
degenerate flavours. Similarly, we can take a fourth root to obtain a single
quark theory, which we take to be the strange quark contribution. For such a
2+1 theory, the fermionic action reads \[\fermionsuper:S_f = \bar\psi
\M(\fermionsuper:m_f =\bar m)^{-1/2} \psi + \bar\chi \M(\fermionsuper:m_f =
\fermionsuper:m_s)^{-1/4} \chi,\] where \(\bar m = (\fermionsuper:m_u +
\fermionsuper:m_d)/2\). With such an action, we can use RHMC as described above
with the two fields \(\psi\) and \(\chi\). The eigenspectrum of the {\asqtad}
operator is bounded from below by \(\fermionsuper:m_f\), so it trivial to
ensure that the rational approximation encompasses all of the spectrum.

A complexity does arise when we consider the calculation of the force
contribution when integrating Hamilton's equations. When performing
\asqtad-type simulations using the R algorithm at typical quark masses, the
computational cost is split roughly equally between the matrix inversion and
the cost of the derivative of the matrix with respect to the gauge field. If we
were to proceed na\"{\i}vely using the same formulation, we would be evaluating
this derivative \(\bar n +\fermionsuper:n_s\) times (where \(\bar n\) and
\(\fermionsuper:n_s\) are the degrees of the MD rational approximations for the
light and strange quark contributions, of \(O(10)\) for typical light masses),
since each term in the partial fraction expansion requires a different field
with a different shift. This would lead to an algorithm approximately
\(\frac12(\bar n+\fermionsuper:n_s)\) more expensive than the R algorithm.

The solution lies in how we calculate this derivative. The {\asqtad} force term
is composed of terms like \(U\ldots U X X^\dagger U\ldots U\), where \(X =
\M^{-1}\phi\). When there are just one or two pseudofermion fields \(\phi\), it
is most efficient to compute the \(U\ldots UX\) products first and then
evaluate the outer product: indeed, this is how the R algorithm is
implemented. However, it is crucial that the shift dependence is only present
in the vectors which appear in the force term. This suggests that for the RHMC
force term, where there are many vectors \(X_i\), if we perform the link matrix
multiplication first, which is shift independent, and then include the outer
product for each partial fraction contribution, we will vastly reduce the
number of operations performed, i.e.,
\begin{eqnarray*}
&& \sum_{i=1}^n (U\ldots UX_i\ ) (X_i^\dagger U\ldots U ) = \\
&&\qquad = U\ldots U \Bigl(\sum_{i=1}^n X_iX_i^\dagger\Bigr) U\ldots U
\end{eqnarray*}

The drawback of this approach is that we will still be doing more operations
than if we were performing the R algorithm calculation. It turns out that the
operations required scale with volume \(V\) as \((782,424 + 720n)V\), where
\(n\) is the degree of the approximation, compared to \(196,920V\) for the R
algorithm. This would imply an approximate four-fold overhead, but since the
only mass dependence in the derivative appears in the vectors, we can combine
the calculation of the derivative for the light and strange contributions, thus
reducing the overhead by about a factor of two. Also, for the 2+1 case, the R
algorithm requires that the {\asqtad} operator be constructed three times
within a single MD step (once for each heatbath, and once for the inversion),
compared to only once for inversion with RHMC. Taking all of the above into
account, we can see that the R algorithm scales for 2+1 like \(828,288V\), so
the R algorithm is actually more expensive.

The current implementation of RHMC for {\asqtad} fermions on the {\qcdoc} is
very competitive with the R algorithm implementation. Their respective
efficiencies are very similar at around \(36\%\) of peak performance. It is
difficult to compare the two algorithms since the R algorithm is an inexact
algorithm, and strictly requires an extrapolation to zero step size.

\section{Domain Wall Simulations}

The case of Furman--Shamir domain wall simulations is quite different to
{\asqtad} type simulations. We now have a 2+1 action as follows
\[\fermionsuper:S_f = \bar\psi
\frac{\M(\fermionsuper:m_f=1)}{\M(\fermionsuper:m_f=\bar m)} \psi + \bar\chi\;
\sqrt{\frac{\M(\fermionsuper:m_f=1)}{\M(\fermionsuper:m_f=\fermionsuper:m_s)}}\;
\chi.\] For the two flavour contribution we can use conventional HMC evolution,
the complication is only present for the strange quark contribution. We
represent the square root which appears in the action by a rational
approximation and proceed using conventional HMC for the degenerate light
contribution and RHMC for the strange contribution. Unfortunately, it is not
quite so simple because of the inclusion of the Pauli--Villars field in the
numerator in the square root. Since \(\fermionsuper:m_f\) does not appear as a
multiple of the identity in the domain wall fermion matrix, it is not possible
to write a rational approximation as a function of this ratio in terms of
shifted matrices. This does not prevent evaluation of such an approximation,
but it does preclude using a multi-shift solver for the evaluation, rendering
the formulation expensive.

The solution to this problem is to split this ratio into separate
fields. The action is then written as
\begin{eqnarray*}
 && \fermionsuper:S_f = \bar\psi
  \frac{\M(\fermionsuper:m_f=1)}{\M(\fermionsuper:m_f=\bar m)} \,\psi + \\ &&
  \quad {} + \bar\phi \M(\fermionsuper:m_f=1)^\frac12 \phi + \bar\chi
  \M(\fermionsuper:m_f=\fermionsuper:m_s)^{-\frac12}\chi,
\end{eqnarray*}
where we have to perform RHMC on the last two fields. Although we are now
including an extra field, with which we have to perform a matrix inversion, the
additional cost is negligible since the mass is \(O(100)\) times greater than
that of the degenerate light pair. Indeed, the cost of including the effects of
the strange quark are small compared to that of the light pair since it too is
relatively heavy. This cost can be reduced by using a Sexton--Weingarten
integration scheme~\cite{sexton92a}, using a smaller step size for the light
quark pair. With such an implementation, the overhead of performing 2+1 over
regular 2 flavour is negligible. It should be noted that we are forced to
perform a matrix inversion to include the bosonic Pauli-Villars field --- a
surprising result. Since the force calculation is trivial for domain wall
fermions, the contribution to the force is performed na\"{\i}vely as the extra
cost is negligible.

We have implemented this algorithm and have tested simple observables to ensure
its correctness. A more systematic study of algorithmic performance using
domain wall fermions shall be forthcoming in a future publication.

\section{Conclusions}

We have demonstrated that RHMC can be used to generate gauge configurations
efficiently when applied to 2+1 {\asqtad} simulations. The na\"{\i}ve cost is
similar to the R algorithm for a given step size, but the latter requires
extrapolation to zero step size.

Initial findings from performing an exact 2+1 domain wall simulation have been
presented. These initial results indicate that RHMC is perfectly suited for
this fermion formulation.

When performing simulations using RHMC we automatically are using
multipseudofermions leading to a Hasenbusch-type force reduction, which can be
utilised to allow an increase in the step size. This accelerates the
performance of RHMC over that of the R algorithm and conventional HMC
significantly. The effect of this acceleration, and how these 2+1 simulations
behave on large volumes and small masses, shall be investigated in initial runs
on the QCDOC.


\begin{thebibliography}{1}

\bibitem{gottlieb87a}
S.~Gottlieb, W.~Liu, D.~Toussaint, R.~L. Renken, and R.~L. Sugar,
\newblock Phys. Rev. {\bf D35}, 2531 (1987).

\bibitem{forcrand96a}
P.~de~Forcrand and T.~Takaishi,
\newblock Fast fermion {Monte} {Carlo},
\newblock  in {\em Lattice '96}, edited by C.~Bernard, M.~Golterman, and
  M.~Ogilvie, Nuclear Physics (Proceedings Supplements) Vol. B53, pp.
  968--970 (1997),
  \href{http://xxx.soton.ac.uk/abs/hep-lat/9608093}{hep-lat/9608093}.

\bibitem{jansen97a}
K.~Jansen and R.~Frezzotti,
\newblock Phys. Lett. {\bf B402}, 328 (1997),
  \href{http://xxx.soton.ac.uk/abs/hep-lat/9702016}{hep-lat/9702016}.

\bibitem{Clark:2003na}
M.~A. Clark and A.~D. Kennedy,
\newblock Nuclear Physics (Proceedings Supplements), {\bf B129},
  850--852 (2004),
  \href{http://xxx.soton.ac.uk/abs/hep-lat/0309084}{hep-lat/0309084}.

\bibitem{duane87a}
S.~Duane, A.~D. Kennedy, B.~J. Pendleton, and D.~Roweth,
\newblock Phys. Lett. {\bf 195B}, 216 (1987).

\bibitem{sexton92a}
J.~C. Sexton and D.~H. Weingarten,
\newblock Nucl. Phys. {\bf B380}, 665 (1992).

\end{thebibliography}
\end{document}